\documentclass{aa}  
\pdfoutput=1
\usepackage[varg]{txfonts}
\usepackage{natbib}
\usepackage{graphicx}
\usepackage{amsmath}	
\usepackage{mathtools}
\usepackage{amssymb}
\usepackage{hyperref}
\usepackage{gensymb}
\hypersetup{
    colorlinks   = true, 
    urlcolor     = blue, 
    linkcolor    = blue, 
    citecolor   = blue 
}
\usepackage{multirow}
\usepackage{tabularx}
\usepackage[group-separator={,}]{siunitx}
\usepackage{bm}
\usepackage{xcolor}
\usepackage[export]{adjustbox}
\usepackage{url}

\usepackage[rightcaption]{sidecap}
\usepackage[switch]{lineno}

\begin{document} 

   \title{Blending effects on shear measurement synergy between \textit{Euclid}-like and LSST-like surveys}
   \authorrunning{S. Zhang et al.}

\author{Shiyang Zhang\inst{1,2}
        \and
        Shun-Sheng Li\inst{1,3}
        \and
        Henk Hoekstra\inst{3}
}

\institute{
Ruhr University Bochum, Faculty of Physics and Astronomy, Astronomical Institute (AIRUB), German Centre for Cosmological Lensing, 44780 Bochum, Germany\\
\email{shiyang@astro.ruhr-uni-bochum.de}
\and
Lorentz Institute for Theoretical Physics, Leiden University,
2333 CA Leiden, The Netherlands
\and
Leiden Observatory, Leiden University, Einsteinweg 55, 2333 CC Leiden, the Netherlands  
}

   \date{Received XXX, XXXX; accepted YYY, YYYY}

  \abstract{Weak gravitational lensing is a powerful probe for constraining cosmological parameters, but its success relies on accurate shear measurements. In this paper, we use image simulations to investigate how a joint analysis of high-resolution space-based and deep ground-based imaging can improve shear estimation. We simulate two scenarios: a grid-based setup, where galaxies are placed on a regular grid to mimic an idealised, blending-free scenario, and a random setup, where galaxies are randomly distributed to capture the impact of blending. Comparing these cases, we find that blending introduces significant biases, particularly in LSST-like data due to its larger point spread function. This highlights the importance of including realistic blending effects when evaluating the performance of joint analyses. Using simulations that account for blending, we find that the most effective catalogue-level synergy is achieved by combining all galaxies detected in either survey. This approach yields an effective galaxy number density of $44.08~\rm arcmin^{-2}$ over the magnitude range of 20.0 to 27.5, compared to $39.17~\rm arcmin^{-2}$ for LSST-like data alone and $30.31~\rm arcmin^{-2}$ for \textit{Euclid}-like data alone. Restricting the analysis to only the overlapping sources detected in both surveys results in a gain of ${\sim}12\%$ compared to using either survey alone. While this joint-object approach is suboptimal at the catalogue level, it may become more effective in pixel-level analyses, where a joint fit to individual galaxy shapes can better leverage the complementary strengths of both data sets. Studying such pixel-level combinations, with realistic blending effects properly accounted for, remains a promising direction for future work.}
    \maketitle

\section{Introduction}
\label{Sec:Intro}

Weak gravitational lensing, the small but coherent distortions in the images of distant objects introduced by the gravitational fields of intervening structures, is a powerful tool for studying cosmology (see \citealt{Bartelmann2001PhR...340..291B}, for a review). 
Weak lensing provides an unbiased probe of the projected matter distribution in the Universe, allowing for constraints on key cosmological parameters (see e.g., \citealt{hoekstra2008weak,Kilbinger2015RPPh...78h6901K}, for reviews).

Recent weak lensing surveys, such as the Kilo-Degree Survey (KiDS; \citealt{de2013kilo}), the Dark Energy Survey (DES; \citealt{dark2016dark}), and the Hyper Suprime-Cam survey (HSC; \citealt{aihara2018first}) delivered some of the tightest cosmological constraints on the clumpiness of matter in the local Universe, quantified by the parameter $S_8 \equiv \sigma_8 \sqrt{\Omega_m/0.3}$, where $\sigma_8$ denotes the standard deviation of linear matter density fluctuations in a sphere of $8h^{-1}~\rm Mpc$ radius, and $\Omega_m$ denotes the total matter density parameter. Notably, the $S_8$ values inferred from these surveys (KiDS, \citealt{asgari2021kids}; DES, \citealt{amon2022dark, secco2022dark}; and HSC, \citealt{dalal2023hyper,li2023hyper}) are consistently lower than those predicted by {\it Planck} cosmic microwave background (CMB) observations \citep{aghanim2020planck} at a modest level of $1-3\sigma$, and its origin is still not well understood. 

Interestingly, the most recent analysis from the complete KiDS-Legacy survey reports results in closer agreement with \textit{Planck} results, showing a difference of only $0.73\sigma$ \citep{wright2025kidslegacycosmologicalconstraintscosmic}, suggesting that this apparent tension may largely attribute to systematic uncertainties. Continued efforts to improve measurement precision and reduce systematics will be essential for clarifying the level of consistency between weak lensing and early Universe constraints.

The next generation of weak lensing surveys will reduce statistical uncertainties by an order of magnitude, thanks to deeper observations with larger sky coverage. The Rubin Observatory Legacy Survey of Space and Time~(LSST; \citealt{Ivezic2019ApJ...873..111I})\footnote{\url{https://www.lsst.org/}} will repeatedly survey the southern sky from the ground, while \textit{Euclid}~\citep{Laureijs2011arXiv1110.3193L, mellier2024}\footnote{\url{https://sci.esa.int/web/euclid/}} and the \textit{Nancy Grace Roman} Space Telescope~\citep{Spergel2015arXiv150303757S}\footnote{\url{https://roman.gsfc.nasa.gov/}} will obtain high-quality data from space. LSST will provide extraordinarily deep imaging, but with a resolution limited by atmospheric turbulence, or seeing. In contrast, \textit{Euclid} will achieve diffraction-limited resolution in its imaging data, though the survey depth is limited by its smaller telescope diameter and observing time.

These surveys will offer an unprecedented statistical constraining power, with the potential to significantly advance our understanding of the Universe. However, realising this potential requires careful mitigation of systematic biases. One of the crucial steps in weak lensing analyses is the accurate measurement of cosmic shear, which characterises the coherent distortion induced by gravitational lensing. Cosmic shear is not directly observable but is statistically inferred from the observed shapes of galaxies. These observed shapes are themselves affected by various instrumental effects. The dominant sources of these distortions include image noise, optical blurring from the telescope, and—specifically for ground-based observations—atmospheric turbulence (see e.g., \citealt{Mandelbaum2018ARA&A..56..393M} for a review). The point-spread function (PSF), which arises from the blurring effects of the telescope optics, detector response, and atmospheric turbulence, circularises the images, reducing the measured shape correlations, and can also introduce preferred orientations, thereby biasing the measured lensing signal (e.g. \citealp{paulin2008point, massey2013origins, melchior2012means, refregier2012noise}). To account for these biases, we require realistic image simulations that replicate both the observing conditions and galaxy properties relevant to weak lensing analyses \citep[e.g.][]{Hoekstra2015MNRAS449685H, Hoekstra2017MNRAS4683295H}. Accordingly, current state-of-the-art image simulations (e.g., \citealp{kannawadi2019towards,Li2022arXiv221007163L, MacCrann2022MNRAS5093371M}) fine-tuned the input parameters to closely reflect real survey environments, reducing residual systematic biases to acceptable levels.

Image simulations can also be used to explore synergies between data sets. The synergy between a LSST-like ground-based survey and \textit{Euclid}-like space-based survey is particularly promising. For instance, the deep multi-band LSST photometry will improve the photometric redshift estimates \citep{graham2020photometric} for \textit{Euclid}, while the sharp \textit{Euclid} images can help mitigate the impact of blended sources. \citet{Rhodes2017ApJS..233...21R} outlined several scientific benefits from a co-processing of LSST and \textit{Euclid} measurements, including improved constraints on cosmological parameters and applications in
galaxy evolution, transient objects detection, solar system science, and galaxy cluster studies. To maximise the scientific return from these missions, a joint effort was launched to explore the prospects of `derived data products' produced by jointly processing the LSST and \textit{Euclid} data sets \citep{Guy2022zndo...5836022G}. 

There are two main strategies for combining data from different surveys: pixel-level and catalogue-level combinations. Pixel-level combination involves jointly processing the imaging data from different surveys, while catalogue-level combination merges measurements after each survey has independently processed its own imaging data. While pixel-level combination can yield better results, it is computationally intensive and challenging to implement. Thus, assessing the benefits of catalogue-level combinations provides a practical starting point. 

This topic was previously studied by \citet{Schuhmann2019arXiv190108586S}, who simulated LSST-like and \textit{Euclid}-like images of isolated galaxies. Their study demonstrated a clear benefit from combining the data sets, and reported only a modest difference (${\sim}5\%$) in the effective number density of galaxies usable for lensing studies between pixel- and catalogue-level combinations. However, by only simulating isolated galaxies, their results did not account for the effects of blending, where light from nearby sources overlaps. This issue is particularly severe in surveys with large PSFs or in deep fields with high source density.

Blending causes several challenges: it biases source detection (e.g., \citealt{Hoekstra2021AA...646A.124H}), affects the selection function of the source sample (e.g., \citealt{Hartlap2011AA528A51H,Chang2013MNRAS4342121C}), and degrades shear measurement accuracy (e.g., \citealt{Hoekstra2015MNRAS449685H,Hoekstra2017MNRAS4683295H, Mandelbaum2018MNRAS4813170M}). Additionally, blending can introduce redshift-dependent biases (e.g., \citealt{MacCrann2022MNRAS5093371M, Li2022arXiv221007163L}) and increased ellipticity dispersion (e.g. \citealp{dawson2015ellipticity}), especially in crowded regions such as galaxy clusters \citep{Martinet2019AA627A59E}. Combining \textit{Euclid} data with LSST analyses can help mitigate blending effects by leveraging  \textit{Euclid}'s superior resolution to deblend, thereby improving shear measurements. Similarly, the greater depth of LSST can improve \textit{Euclid} analyses by detecting faint sources. The complementary strengths of both surveys enhance joint analyses, and enable more robust weak lensing analyses.

In this paper, we extend the study of \citet{Schuhmann2019arXiv190108586S} by accounting for the blending of galaxies. Specifically, we focus on the synergy between LSST-like and \textit{Euclid}-like surveys at the catalogue level, leaving the more complex joint-pixel analysis for future work. 
We examine two catalogue-level combination strategies: the joint case, which includes only sources detected in both surveys; and the combined case, which includes all detected sources from either survey. These combination strategies allow us to quantify how catalogue-level synergy improves the effective number density and shear measurements.

To quantify the impact of blending, 
we simulate LSST-like and \textit{Euclid}-like images with two different scenarios: grid-based and randomly placed configurations. The grid-based simulations place galaxies on a regular grid with fixed separations, providing an idealised setup free from blending effects. In contrast, the randomly placed simulations place galaxies randomly in the images, allowing natural overlaps and thereby capturing blending effects expected in real observations. By comparing these two scenarios, we assess the impact of blending on shear measurements. 
Notably, our simulations do not include galaxy clustering, which will further increase the impact of the blending effect \citep{Martinet2019AA627A59E,Li2022arXiv221007163L}. Therefore, our estimates of blending effects are conservative, and future studies incorporating realistic galaxy clustering are necessary for a more comprehensive assessment.

The remainder of this paper is organised as follows. In Sect.~\ref{Sec:ImaSim}, we introduce the \textit{Euclid}-like and LSST-like image simulations. The impact of blending on source detection is examined in Sect.~\ref{Sec:Detec}. In Sect.~\ref{Sec:Shape}, we describe the shape measurement process and shear calibration for both surveys. We detail the impact of blending on shape noise estimation in Sect.~\ref{Sec:SN}. Section~\ref{Sec:ShearComp} outlines our approach for quantifying the synergy. The results are presented in Sect.~\ref{Sec:Results}, and we conclude in Sect.~\ref{Sec:Conclusion}.

\section{Image simulations}
\label{Sec:ImaSim}

\begin{table*}
\renewcommand{\arraystretch}{1.5}
\caption{Overview of simulation setups.}
\label{Tab: setup}
\centering
\begin{tabular}{l c c c c}
\hline\hline
\multicolumn{1}{l}{} & \multicolumn{2}{c}{LSST-like}       & \multicolumn{2}{c}{\textit{Euclid}-like}                 \\ \hline
Pixel scale          & \multicolumn{2}{c}{$0\farcs2$}           & \multicolumn{2}{c}{$0\farcs1$}                         \\
Effective background noise & \multicolumn{2}{c}{A $5\sigma$ point source depth of 26.9} & \multicolumn{2}{c}{A $5\sigma$ point source depth of 26.3} \\
PSF profile &
  \multicolumn{2}{p{0.3\textwidth}}{Circular Moffat profile with parameters drawn from the KiDS survey} &
  \multicolumn{2}{p{0.3\textwidth}}{Circular Airy profile for a telescope with a diameter of 1.2m and an obscuration of 0.3} \\ \hline
Galaxy positions & Grid       & Random         & Grid      & \multicolumn{1}{c}{Random} \\
Magnitude cut        & {[}20.0, 25.2{]} & {[}20.0, 27.5{]} & {[}20.0, 25.2{]} & {[}20.0, 27.5{]}             \\ \hline
Effective simulated area  & \multicolumn{4}{c}{$100~{\rm deg}^2$}     \\ 
Input shear $(g_1, g_2)$   & \multicolumn{4}{c}{($+0.0283, +0.0283$), ($+0.0283, -0.0283$), ($-0.0283, +0.0283$), ($-0.0283, -0.0283$)}                                        \\
Shape noise cancellation & \multicolumn{4}{c}{Each image is paired with galaxies rotated by 90 degrees}  \\ \hline
\end{tabular}
\end{table*}

To explore the benefits of a joint analysis of \textit{Euclid} and LSST data for weak lensing studies, we use simulated images  
generated from the \texttt{MultiBand\_ImSim} pipeline\footnote{\url{https://github.com/KiDS-WL/MultiBand_ImSim}}~\citep{Li2022arXiv221007163L}, which is based on the public \textsc{GalSim} package\footnote{\url{https://github.com/GalSim-developers/GalSim}}~\citep{rowe2015galsim}. 
The input galaxy population is drawn from the \textsc{surfs}-based KiDS-Legacy-Like Simulation (SKiLLS), a galaxy mock catalogue that combines cosmological simulations with high-quality imaging observations to achieve realistic galaxy properties, including multi-band photometry, galaxy morphology, and their correlations~\citep{Li2022arXiv221007163L}. 
Its backbone cosmological simulations are the Synthetic UniveRses For Surveys (\textsc{surfs}) $N$-body simulations~\citep{Elahi2018MNRAS} with the galaxy properties created from an open-source semi-analytic model named \textsc{Shark}\footnote{\url{https://github.com/ICRAR/shark}}~\citep{Lagos2018MNRAS}. 
The observed galaxy morphologies are modelled with S\'ersic profiles \citep{sersic1963BAAA641S} using three parameters: the effective radius, which determines the size of galaxies; the S\'ersic index, which describes the concentration of the brightness distribution; and the axis ratio of the projected profiles. These structural parameters are derived from high-quality imaging observations obtained with the Advanced Camera for Surveys instrument on the \textit{Hubble} Space Telescope~\citep{Griffith2012ApJS}. We refer to \citet{Li2022arXiv221007163L} for details on the learning algorithm and validation.

The 10-year LSST survey is expected to achieve a nominal $5\sigma$ point source depth of 26.88 mag in the $r$-band~\citep{bianco2021optimization}, which is close to the limiting magnitude of 27 in the original SKiLLS simulations. We modelled background noise as Gaussian fluctuations and ignored the Poisson photon noise from individual galaxies, which is a valid approximation as we focused on galaxies with modest signal-to-noise ratio (S/N) for which the background noise dominates. We adopted the AB magnitude system throughout and configured SExtractor with a magnitude zero-point of 30, so that the measured flux uncertainties directly reflect the imposed Gaussian background and depth. \citet{Martinet2019AA627A59E} showed that undetected faint galaxies, especially when clustered around detected sources, can introduce non-negligible biases in shear measurements. Therefore, to better account for the influence of such undetected objects, we extend the original SKiLLS catalogue to a depth of 30 mag in the $r$-band. To do so, we divide the \textsc{surfs}-\textsc{Shark} galaxies into magnitude bins with a width of 0.1 mag. In each bin, we iteratively select galaxies until the count matches the predictions from the analytical fitting function by \citet{fenech2017calibration}. The repetition factor, representing the incompleteness of the baseline \textsc{surfs}-\textsc{Shark} catalogue at the faint end, increases from 2 in the [27.6, 27.7] bin to 130 in the [29.9, 30.0] bin. The morphologies of these added faint galaxies are learned from the \textit{Hubble} Ultra Deep Field observations~\citep{Coe2006AJ....132..926C} using the same learning algorithm as in \citet{Li2022arXiv221007163L}. We note that these faint objects mainly introduce correlated background noise, and their detailed properties are less crucial. 

To generate \textit{Euclid}-like images, we follow \cite{Hoekstra2017MNRAS4683295H} and approximate the \textit{Euclid} PSF by a circular Airy profile for a telescope with a diameter of $1.2$m and an obscuration of $0.3$ at a reference wavelength of $800$nm, which corresponds to a diffraction-limited full-width-half-maximum (FWHM) of ${\sim}0\farcs16$. 
The pixel size is set to $0\farcs1$ per pixel  \citep{Cropper2018SPIE10698E..28C}. The background noise of the image is assumed to be Gaussian and tuned to mimic the depth of four coadded \textit{Euclid} exposures~\citep{ilbert2021euclid}. The resulting mock images have a $10\sigma$ extended source limiting magnitude of $24.5$, equivalent to a $5\sigma$ point source limiting magnitude of $26.3$.\footnote{\cite{mellier2024} report 
a $5\sigma$ point source limiting magnitude of $m_{\rm AB}=26.7$ for the actual \textit{Euclid} data.}

To generate the LSST-like images, we assume that the atmospheric conditions are similar to those experienced by KiDS. This allows us to draw a realistic distribution of PSF properties from these data. 
Specifically, we use stacked $r$-band images from the fourth public data release of KiDS~\citep{kuijken2019fourth} and fit Moffat profiles to bright stars (see \citealt{Li2022arXiv221007163L} for details). 
We adopt a constant PSF with FWHM of $0\farcs7$, matching the average seeing measured in stacked $r$-band exposures from the fourth public KiDS data release~\citep{kuijken2019fourth} and fit Moffat profiles to bright stars (see \citealt{Li2022arXiv221007163L} for details). We ignore PSF anisotropy for simplicity.
The pixel size is set to $0\farcs2$ per pixel \citep{Ivezic2019ApJ...873..111I}. The background noise of the image is assumed to be Gaussian and adjusted to match the LSST $r$-band depth after the 10-year survey, with a $5\sigma$ point source depth of $26.88$ mag \citep{bianco2021optimization}.

We keep the image layout simple and use the same field-of-view, covering an area of $0.5\degree\times 0.5\degree$ for both the \textit{Euclid}-like and LSST-like images, so that the two surveys cover the same input galaxy population. This setup corresponds to a joint analysis of the \textit{Euclid} and LSST surveys in their overlapping regions. All images are simulated using galaxy properties in the $r$-band, ignoring any colour-related galaxy property variations. The observational conditions are matched to \textit{Euclid}’s VIS and LSST $r$ filter, respectively.

\begin{figure*}
\centering
\includegraphics[width=17cm]{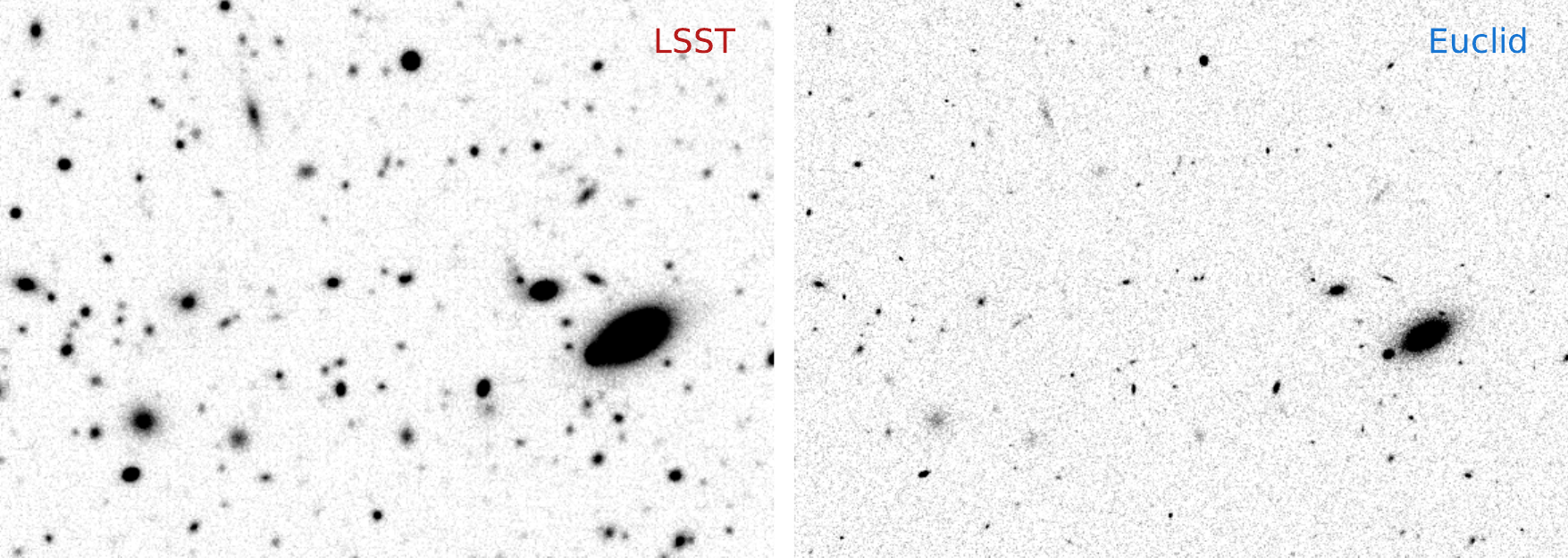}
\caption{Cut-outs of a $1\farcm8 \times 1\farcm2$ region for a 10-year LSST-like survey with randomly placed sources (left) and the corresponding \textit{Euclid}-like image (right). The PSF of the LSST-like image has FWHM of $0\farcs7$, while that of the 
\textit{Euclid}-like image is $0\farcs16$. The effective background noise corresponds to a equivalent surface brightness of 27.50 $\rm mag~arcsec^{-2}$ for the LSST-like images and 24.71 $\rm mag~arcsec^{-2}$ for the \textit{Euclid}-like images. Detailed information on the image simulation setup is provided in Table~\ref{Tab: setup}.}
\label{Fig: lsst_euclid}
\end{figure*}

We generate four sets of simulations using the same input galaxy catalogue: grid-based \textit{Euclid}-like images, grid-based LSST-like images, randomly placed \textit{Euclid}-like images, and randomly placed LSST-like images. For the grid-based simulations, we place the galaxies on a grid with a constant separation of $18\arcsec$ to ensure a blending-free scenario. In the case of the randomly placed galaxies, the profiles are allowed to overlap. By comparing these two different setups, we can quantify the impact of blending. Each set of simulations contains 50 realisations initialised with different random seeds that specify the Gaussian background noise and galaxy positions in the cases of randomly placed galaxies. To estimate the shear bias, we repeat each image with four different input shear values: $(0.0283, 0.0283)$, $(0.0283, -0.0283)$, $(-0.0283, 0.0283)$, $(-0.0283, -0.0283)$. To reduce the shape noise, we pair each image with a counterpart, where the galaxies are rotated by $90$ degrees, while all the other setups remain the same. Example images for randomly placed cases are shown in Fig.~\ref{Fig: lsst_euclid}, and the image simulation setup is summarised in Table~\ref{Tab: setup}. For simplicity, in the following discussion, we refer to grid-based simulations as `grid' and randomly placed simulations as `random'.

\section{Blending on object detection}
\label{Sec:Detec}

The first step in the lensing analysis is source detection,
for which we use \textsc{SExtractor} \citep{bertin1996sextractor}; the relevant configuration parameters are listed in Table \ref{tab: sex_setup}. 
After detection, we cross-matched each input galaxy to the SExtractor catalogue within $3\arcsec$. If multiple galaxies lie within this radius, the galaxy is assigned to the detection whose \texttt{MAG\_AUTO} is closest to its input $r$-magnitude. Hereafter, input magnitude refers to the $r$-magnitude of this matched input galaxy.
Shear measurement biases are introduced already during source detection \citep{fenech2017calibration, kannawadi2019towards}, and blending further exacerbates these biases \citep{Hoekstra2021AA...646A.124H}. To compare the impact of blending on source detection, we create a joint detection catalogue by cross-matching the detected galaxies from the \textit{Euclid}-like simulations and the LSST-like simulations, preserving only objects detected in both simulations. We find that the source number density of this cross-matched detection catalogue is $32.25 \rm~ arcmin^{-2}$ for the random simulations, compared to an equivalent number density of $33.95 \rm ~arcmin^{-2}$ for the grid simulations within the same input magnitude range of $\left[20.0, 25.2\right]$.

\begin{table}[]
\centering
\caption{Relevant SExtractor setup parameters for both survey.}
\label{tab: sex_setup}
\resizebox{\columnwidth}{!}{
\begin{tabular}{lll}
\hline
\hline
                             & LSST-like     & $Euclid$-like     \\\hline
\texttt{DETECT\_MINAREA}  & 4            & 16           \\
\texttt{DETECT\_THRESH}   & 1.5          & 1.5          \\
\texttt{FILTER\_NAME}     & \texttt{default.conv} & \texttt{default.conv} \\
\texttt{DEBLEND\_NTHRESH} & 32           & 32           \\
\texttt{DEBLEND\_MINCONT} & 0.001        & 0.001        \\
\texttt{CLEAN\_PARAM}     & 1.0          & 1.0          \\
\texttt{BACK\_SIZE}       & 64           & 256          \\
\texttt{BACK\_FILTERSIZE} & 3            & 3           \\ \hline

\end{tabular}
}
\end{table}

\begin{figure}
\centering
\includegraphics[width=\hsize]{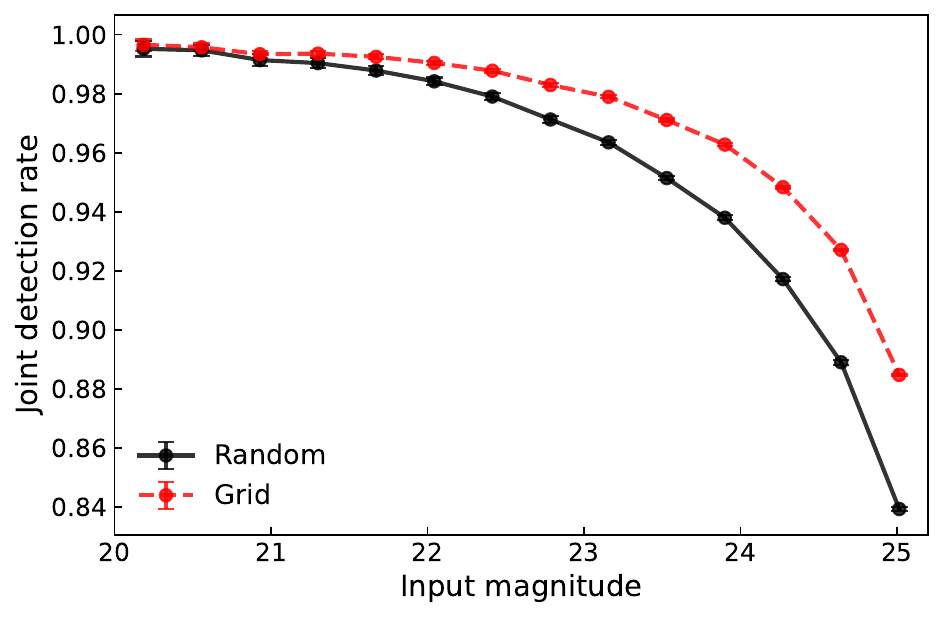}
\caption{The fraction of jointly detected galaxies as a function of the input $r$-band magnitude. The black solid line represents the results from the random simulations, while the red dashed line corresponds to the grid simulations.}
\label{Fig: detec_both}
\end{figure}

Figure~\ref{Fig: detec_both} shows the fraction of galaxies in the joint detection catalogues as a function of input magnitude. To provide a direct comparison, we use input magnitudes throughout the analysis to isolate the effects of blending from other observational systematics. The red dashed line represents the joint detection rate from the grid simulations, while the black solid line corresponds to those from the simulations with randomly placed galaxies. The fraction is lower for the latter, demonstrating the non-negligible impact from the blending effects, and thus highlighting the importance of accounting for blending effects when evaluating the synergy between \textit{Euclid}-like and LSST-like surveys. 

\begin{figure}
\centering
\includegraphics[width=\hsize]{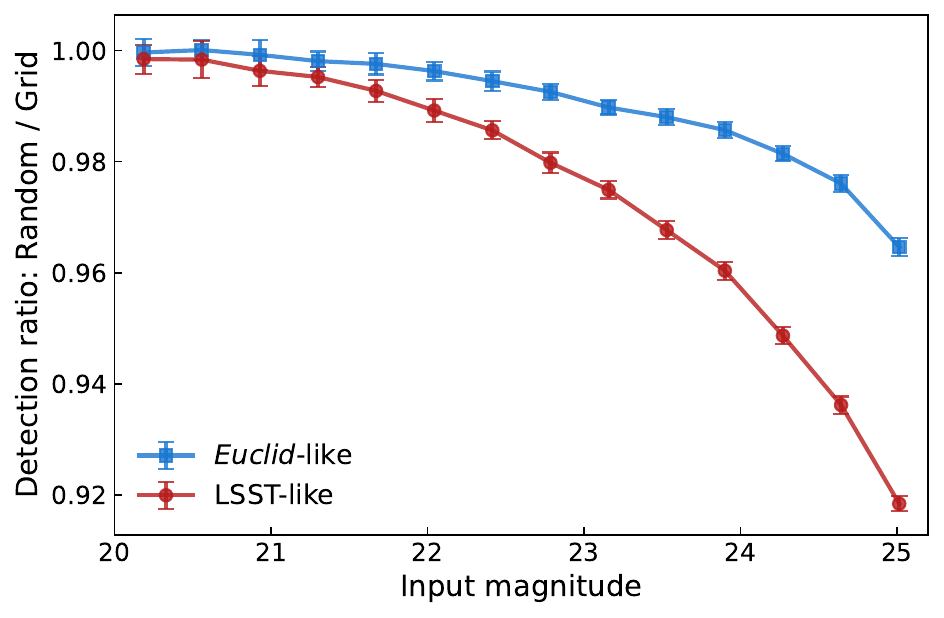}
\caption{The ratio of detected number densities between random and grid cases for each survey (blue: \textit{Euclid}-like; red: LSST-like). For reference, for galaxies with an input magnitude of 25, the mean S/N, computed as the average of \texttt{FLUX\_AUTO / FLUXERR\_AUTO}, is 7.13 for the \textit{Euclid}-like survey and 16.20 for the LSST-like survey when galaxies are placed randomly.}
\label{Fig: detec_ratio}
\end{figure}

To further investigate the blending effect, Figure~\ref{Fig: detec_ratio} shows the relative detection efficiency between the random and grid setups for individual surveys as a function of the $r$-band input magnitude. In both cases, the random configuration yields fewer detected galaxies, and the detection ratio is decreasing toward fainter magnitudes. The change, however, is higher for the LSST-like setup (red line), compared to the \textit{Euclid}-like case, indicating that blending affects the LSST-like survey more significantly. Quantitatively, within the input magnitude range of $\left[20.0, 25.2\right]$, the detected number density in the LSST-like simulations decreases by ${\sim}5\%$, from $38.27 ~\rm arcmin^{-2}$ in the grid setup to $36.35 \rm ~arcmin^{-2}$ in the random setup. This effect is smaller in the \textit{Euclid}-like simulations, where the number density drops by only ${\sim}2\%$, from $34.00$ to $33.38~\mathrm{arcmin^{-2}}$.

\begin{figure*}
\centering
\includegraphics[width=\hsize]{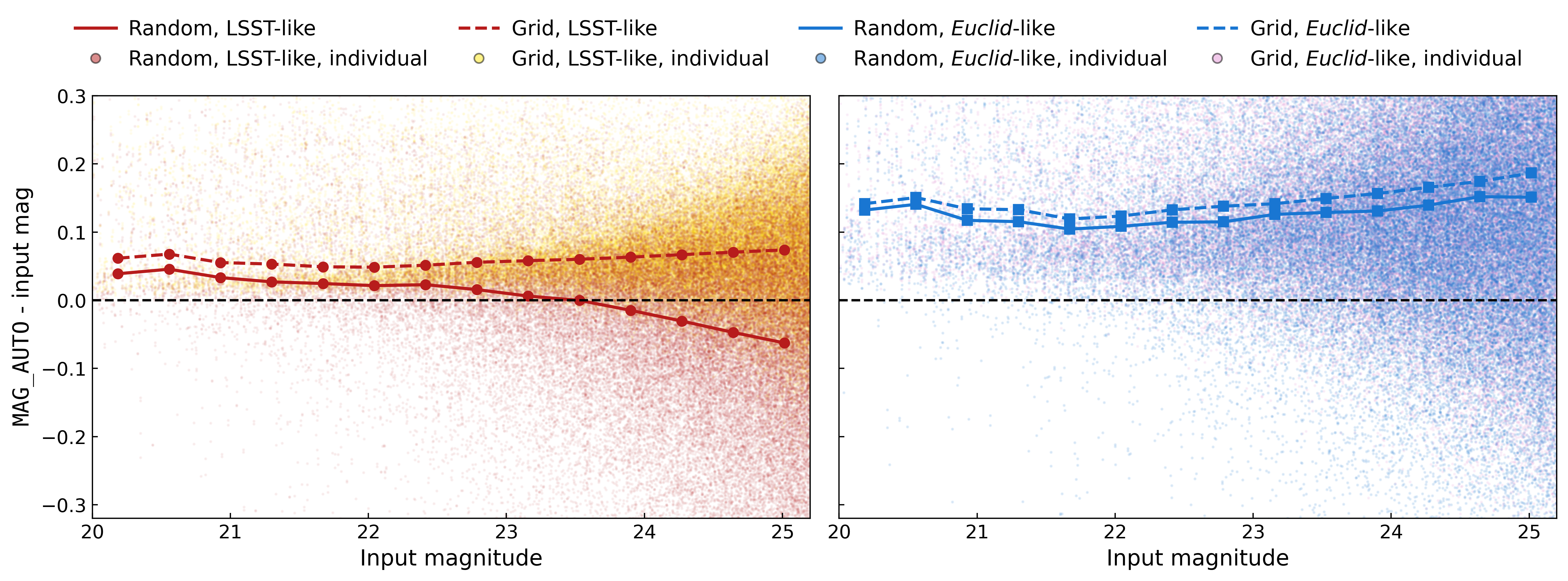}
\caption{The difference between the measured magnitude and the input magnitude as a function of the input magnitude. 
Background points show individual galaxies for the all cases. The solid lines trace the median values of the difference for each magnitude bin in the random cases and the dashed lines show the median for the grid cases.}
\label{Fig: mag_corr}
\end{figure*}

Blended sources that are not individually detected contribute additional flux to neighbouring objects, affecting their observed magnitudes \citep[see figure~2 in][]{Hoekstra2021AA...646A.124H}. To investigate this, we compare the values of \texttt{MAG\_AUTO} reported by \textsc{SExtractor} to the input magnitudes of the sources, where \texttt{MAG\_AUTO} provides an estimate of the `total magnitude' by integrating pixel values within an adaptive aperture defined using the algorithm proposed by \citet{Kron1980ApJS...43..305K}. Figure~\ref{Fig: mag_corr} shows the median of the difference between the measured and input magnitudes as a function of the input magnitude, with the distribution of the galaxies  plotted as points in the background. For the LSST-like grid simulations (dashed red lines in the left panel), we observe a median offset that is nearly independent of magnitude, and the main locus lies slightly above zero, which we attribute to Kron aperture flux loss caused by the extended wings of the Moffat PSF combined with pixel noise. \textit{Euclid}-like simulations (dashed blue lines in the right panel) show a larger median offset, which is likely caused by the sharper core and extended wings of the diffraction-limited Airy PSF we used in the simulation. Also in this case, the median offset does not depend much on the magnitude of the sources. The situation changes when we include blending, especially for the LSST-like setup (solid red line). For the faintest magnitudes, the extra flux from blended sources actually leads to an underestimate of \texttt{MAG\_AUTO} of ${\sim}0.1$ mag, while the impact on the \textit{Euclid}-like (solid blue line) is an overall small shift, with only a slight magnitude dependence. This underestimation of object magnitude will affect the S/N reported by the galaxy shape measurement algorithm, complicating the relationship between shear measurement properties and observed object brightness. This further underscores the importance of accounting for blending in simulations used for weak lensing studies.

In the remainder of the paper, we explore further how blending reduces the effective number density of sources. For the shape measurement step, the change in magnitude owing to blending is minimal, but we note that this may not be the case for the determination of the source redshift distribution, $n(z)$. Specifically, spectroscopic redshift surveys used for calibration may systematically exclude blended sources. Consequently, a selection based on apparent magnitude in the calibration sample may not correspond to the same lensing source sample identified in deep ground-based data. We leave this issue for future investigation.

\section{Shear measurement and calibration}
\label{Sec:Shape}

In this section, we describe our implementation of the shear measurement step. In Sect.~\ref{Sec:ShapeIn} we describe the 
\textsc{ngmix} algorithm (\citealt{sheldon2014implementation}) that we used to estimate the ellipticities of individual galaxies. 
To estimate the shear, these measurements are combined into ensemble averages using a weighting scheme described in Sect.~\ref{Sec:ShearWeight}.
We determine the biases in the shear estimates in Sect.~\ref{Sec:ShearBias}, which is crucial for accurately assessing the performance of each survey and their synergy.

\subsection{Individual galaxy shape measurement}
\label{Sec:ShapeIn}

We use the public \textsc{ngmix} code\footnote{\url{https://github.com/esheldon/ngmix}} (\citealt{sheldon2014implementation}) to estimate the ellipticities for individual galaxies. \textsc{ngmix} employs a Bayesian model-fitting approach to measure the shapes of galaxies, assuming Gaussian mixture models for both galaxies and PSF profiles. This yields an estimate for the galaxy ellipticity, corrected for the blurring by the PSF, given by:
\begin{equation}
    {\bm\epsilon} \equiv \epsilon_1 + i\epsilon_2 = \left(\frac{1-q}{1+q}\right)\exp (2i\phi)~,
\end{equation}
where $q$ and $\phi$ denote the axis ratio, and the position angle of the major axis, respectively.

\textsc{ngmix} has been extensively tested in previous cosmic shear analyses from the Dark Energy Survey~\citep{jarvis2016science,Zuntz2018MNRAS.481.1149Z}. In our analysis, we use uniform priors for the model parameters, ignore the PSF modelling uncertainties, and directly provide the simulation input PSF to \textsc{ngmix}. We note that our setup is idealised compared to real-world data analyses, but we do not expect this to affect our conclusions. The choice of shape measurement algorithm may have an impact, but \textsc{ngmix} is able to return meaningful shapes for a large fraction of the sources. Specifically, we find that ${\sim} 59.22\%$ of sources in the LSST-like grid case and ${\sim} 56.19\%$ in the \textit{Euclid}-like grid case have a S/N in shape measurement greater than 1, where the S/N is defined as the ratio of measured ellipticity to its measurement uncertainty, as reported by \textsc{ngmix}. Therefore, we expect that the results obtained using \textsc{ngmix} are representative among the modern model fitting methods. We account for residual shear biases in Sect.~\ref{Sec:ShearBias}. 

\begin{figure}
\centering
\includegraphics[width=\hsize]{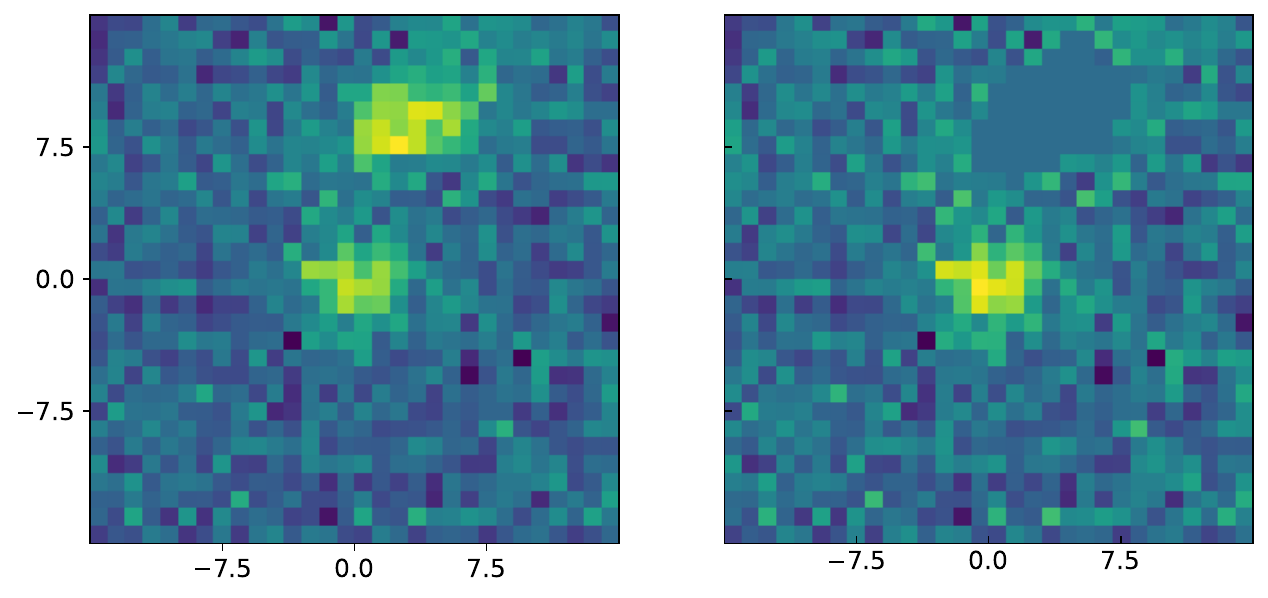}
\caption{An example stamp before (left) and after (right) masking. The neighbouring object is masked by replacing its pixels with zero values.}
\label{Fig: mask}
\end{figure}

We create postage stamps with a size of $6\arcsec {\times} 6\arcsec$ for all detected galaxies as the input to \textsc{ngmix}. The stamp centres are placed at the locations of the targeted objects, as reported by \textsc{SExtractor}. To reduce contamination by light from neighbouring objects, we apply masks to all adjacent detected objects within the target stamps, using the segmentation maps produced by \textsc{SExtractor}. We show in Fig.~\ref{Fig: mask} an example stamp before and after masking. For consistency, this masking process is implemented in both random and grid simulations, although it has no impact on the latter. 

We do not attempt to deblend any unrecognised blends. This is an 
active field of study in itself (see, e.g.~\citealt{Melchior2018AC....24..129M,Reiman2019MNRAS.485.2617R,Arcelin2021MNRAS.500..531A,Buchanan2022ApJ...924...94B,Wang2022PhRvD.106f3023W,Hemmati2022ApJ...941..141H}). Consequently, our results on the impact of blending are somewhat conservative. Nonetheless, they provide a useful baseline, representative of the performance of currently employed shear measurement pipelines.

\subsection{Weighting scheme for shear estimation}
\label{Sec:ShearWeight}

If we assume that galaxies are oriented randomly in the absence of lensing, we can estimate the lensing shear signal $g$ from an ensemble average of galaxies. The intrinsic ellipticity distribution of the sources is the dominant source of uncertainty in shear estimation, especially for bright galaxies. This can, however, depend on galaxy type, brightness, and redshift \citep[e.g.][]{kannawadi2019towards}. Moreover, the precision of the individual measurements varies due to noise and blending. Therefore, we assign weights, $w_i$ to the galaxies to estimate the shear following (e.g.~\citealt{Bartelmann2001PhR...340..291B}):
\begin{equation}
\label{eq:gobs}
    {\bm g}^{\rm obs} = \frac{\sum_i w_i {\bm\epsilon}_i}{\sum_i w_i},
\end{equation}
where the weight $w_i$ depends on two main sources of uncertainty: the intrinsic shape noise of galaxies, and the measurement uncertainties in the ellipticity components. We can calculate a weight for each galaxy measurement following the common inverse-variance approach (\citealt{miller2013bayesian, guinot2022shapepipe}):
\begin{equation}
\label{eq:weight}
w \equiv \frac{1}{2\sigma^2_{\rm in} + \Delta^2_{\epsilon_1}+ \Delta^2_{\epsilon_2}},
\end{equation}
where $\sigma_{\rm in}$ denotes the intrinsic shape noise corresponding to the input ellipticities of the detected galaxies, and $\Delta_{\epsilon}$ represents the ellipticity measurement uncertainties reported by \textsc{ngmix}.

The intrinsic shape noise is defined as the standard deviation per component of the intrinsic ellipticity distribution of detected galaxies. We estimate it directly from the input ellipticities of the detected galaxies in our simulations. This ensures that the estimated intrinsic shape noise accounts for detection selections, while being free from biases in the shape measurements. 

The measurement uncertainty is taken from the estimates reported by \textsc{ngmix}. While \cite{Schuhmann2019arXiv190108586S} found that \textsc{ngmix} underestimates the measurement uncertainties, especially for faint and small galaxies, the resulting bias is small compared to the shape noise from the intrinsic galaxy ellipticities. In practice, such biases in reported uncertainties should still be accounted for when working with observational data. However, in our case, we verified that the weights we define do not show any significant dependence on source ellipticity, indicating that this bias has a negligible effect on our results.

\subsection{Shear measurement biases}
\label{Sec:ShearBias}

The estimates for the ellipticity are biased because of pixel noise and the distortions from the PSF, while blending complicates matters further. Consequently, the shear estimated from the ensemble average of measured ellipticities is also biased, which can be expressed, to first-order, as 
\citep{heymans2006shear}
\begin{equation}
\label{eq:mc}
g^{\rm obs}_{\alpha} = (m_{\alpha}+1)~g^{\rm true}_{\alpha} + c_{\alpha}~,
\end{equation}
where $m_{\alpha}$ and $c_{\alpha}$ are known as the multiplicative bias and additive bias, respectively, and $\alpha \in \{ 1, 2\}$ refers to the two components of the shear. 
For each realisation, we applied four sign combinations of the fiducial input shear $(\pm 0.0283,\pm 0.0283)$ to the intrinsic galaxy shapes before image simulation. We computed the ensemble‐average observed shear $g_{\alpha}^{\rm obs}$ from Eq.~(\ref{eq:gobs}). A linear fit between the observed and input shears yields a best-fit slope corresponding to $m_\alpha + 1$. We report the average multiplicative bias $m_\alpha$ across all 50 realisations.

PSF anisotropy or read-out related detector effects lead to non-vanishing additive bias, where the values of $c_1$ and $c_2$ can differ. As we employ
axisymmetric PSFs to simulate the two surveys, and do not include detector effects, we expect the additive bias to vanish. Indeed, we find that $c_1$ and $c_2$ are consistent with zero for all configurations over the full range of input magnitudes. In contrast, we do not expect the multiplicative bias to vanish, but the values for $m_1$ and $m_2$ should be consistent with one another for our setup. We find this to be the case, and therefore report the amplitude of the multiplicative bias as $m=(m_1+m_2)/2$. 

\begin{figure}
\centering
\includegraphics[width=\hsize]{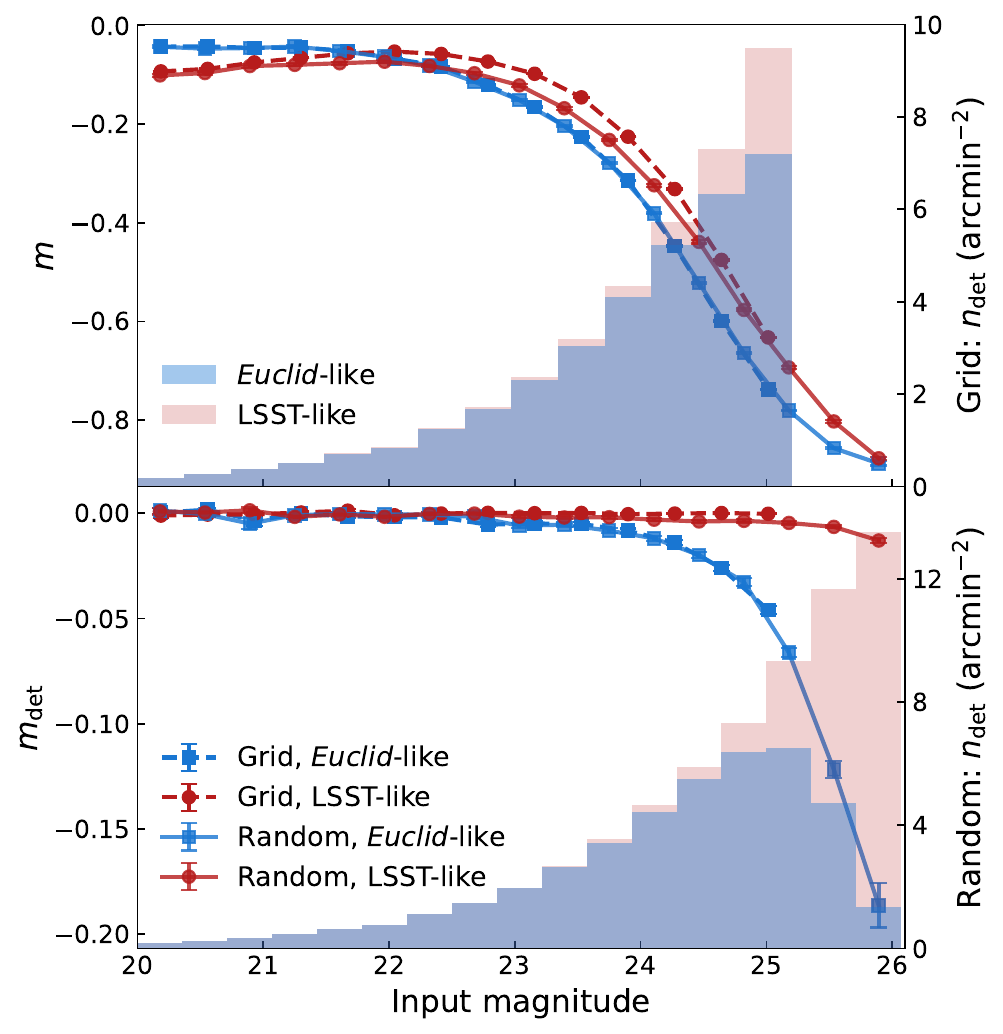}
\caption{Upper panel: amplitude of the multiplicative bias, defined as $m\equiv(m_1+m_2)/2$, as a function of the input $r$-band magnitude of galaxies.  Lower panel: the multiplicative detection bias $m_{\rm det}$ as a function of the input magnitude, with galaxies placed on a grid (dashed) or randomly (solid) for both surveys, also defined as $m_{\rm det}\equiv(m_{\rm det, 1}+m_{\rm det, 2})/2$. The magnitude ranges considered are $\left[20.0, 25.2\right]$ for the grid case and $\left[20.0, 27.5\right]$ for the random case. The histograms illustrate the detected number density of galaxies, with grid case shown in the upper panel and random case in the lower panel. }
\label{Fig: m1_4cases}
\end{figure}

The upper panel of Fig.~\ref{Fig: m1_4cases} shows that the value of multiplicative bias increases rapidly for fainter galaxies and remains nonzero even for bright samples. For the \textit{Euclid}-like simulations (blue lines), the multiplicative biases for the grid (dashed line) and random (solid line) cases are almost indistinguishable. In contrast, the LSST-like simulations (red lines) show a noticeably larger bias in the random case, driven by blending effects \citep[see also][]{Hoekstra2021AA...646A.124H}. While the sharper \textit{Euclid} PSF helps mitigate blending effects, its advantage is offset by higher background noise, especially for faint galaxies with input magnitudes $\gtrsim 22.5$. In this regime, the greater depth of the LSST-like survey yields better performance, resulting in a smaller multiplicative bias at the faint end.

For reference, we also estimate the detection bias, which arises because gravitational lensing conserves surface brightness. When detection algorithms, such as \textsc{SExtractor}, apply surface brightness thresholds and circular kernels, this conservation leads to biased galaxy detections, resulting in an underestimation of the average shear \citep[e.g.][]{Hoekstra2021AA...646A.124H}. Following \citet{fenech2017calibration}, we estimate the detection bias by computing the observed shear from sheared input ellipticities of galaxies $\bm g^{\rm obs}= \sum_i \bm\epsilon^{\rm s}_{i}$,where $\bm \epsilon^{\rm s}$ are the true galaxy ellipticities after applying the input shear and fitting \(g^{\rm obs}_\alpha\) against the known input shear \(g^{\rm true}_\alpha\) as Eq.~(\ref{eq:mc}).

As shown in the lower panel of Fig.~\ref{Fig: m1_4cases}, most of the multiplicative bias originates from the shape measurement rather than the detection process. However, the impact of detection bias differs between the two surveys. The LSST-like survey, with its greater depth, shows negligible detection bias; as illustrated by the red lines, the detection bias remains small and nearly constant across the showed magnitude range. In contrast, the \textit{Euclid}-like survey shows a stronger magnitude dependence. While the detection bias is small within the target magnitude range \citep[$m < 24.5$;][]{Laureijs2011arXiv1110.3193L}, it increases significantly for fainter sources\footnote{In our study, the background noise level for the \textit{Euclid}-like survey was set according to the design requirements from \citet{Laureijs2011arXiv1110.3193L}. The resulting images are shallower than those simulated in \citet{Hoekstra2021AA...646A.124H}, which causes the detection bias to become relevant at comparatively brighter magnitudes.}. This comparison highlights the complementary strengths and limitations of the two surveys: LSST-like data benefit from deeper imaging but suffer more from blending due to their broader PSFs, while \textit{Euclid}-like data offer sharper imaging but are more susceptible to detection bias at faint magnitudes due to their shallower depth. These differences imply the potential benefits of combining both surveys to improve weak lensing analyses.

At the very faint end, the inferred value for $m$ levels off with increasing uncertainties. This is particularly clear for the \textit{Euclid}-like setup. For these faint, noisy sources, we cannot determine the bias robustly. Therefore, to simplify the analysis, we introduce a magnitude cut at the point where $m$ levels off. Specifically, this corresponds to an input $r$-magnitude limit of 26.07 for the \textit{Euclid}-like simulations and 26.79 for the LSST-like simulations. However, we do not expect this to affect our conclusions, because the number density of sources is already low at these magnitudes, as shown by the detected number density in Fig.~\ref{Fig: m1_4cases}. Combined with the lower shear weight, these faint sources carry negligible weight in the analysis.

When comparing the uncertainties in the shear estimates between surveys, it is important to account for the multiplicative shear bias, as it affects how measurement uncertainties should be propagated. Following Eq.~(\ref{eq:mc}), to first-order, we use
\begin{equation}
\label{eq:uncertainty_corrected}
    \sigma_{\epsilon} = \frac{\Delta_{\epsilon}}{1+m}~.
\end{equation}
In the following, we consistently use $\sigma_{\epsilon}$ instead of the raw $\Delta_{\epsilon}$ in the context of comparing between surveys or conducting survey synergy. 

\section{Blending on shape noise estimation}
\label{Sec:SN}

\begin{figure*}
\sidecaption
\includegraphics[width=12cm]{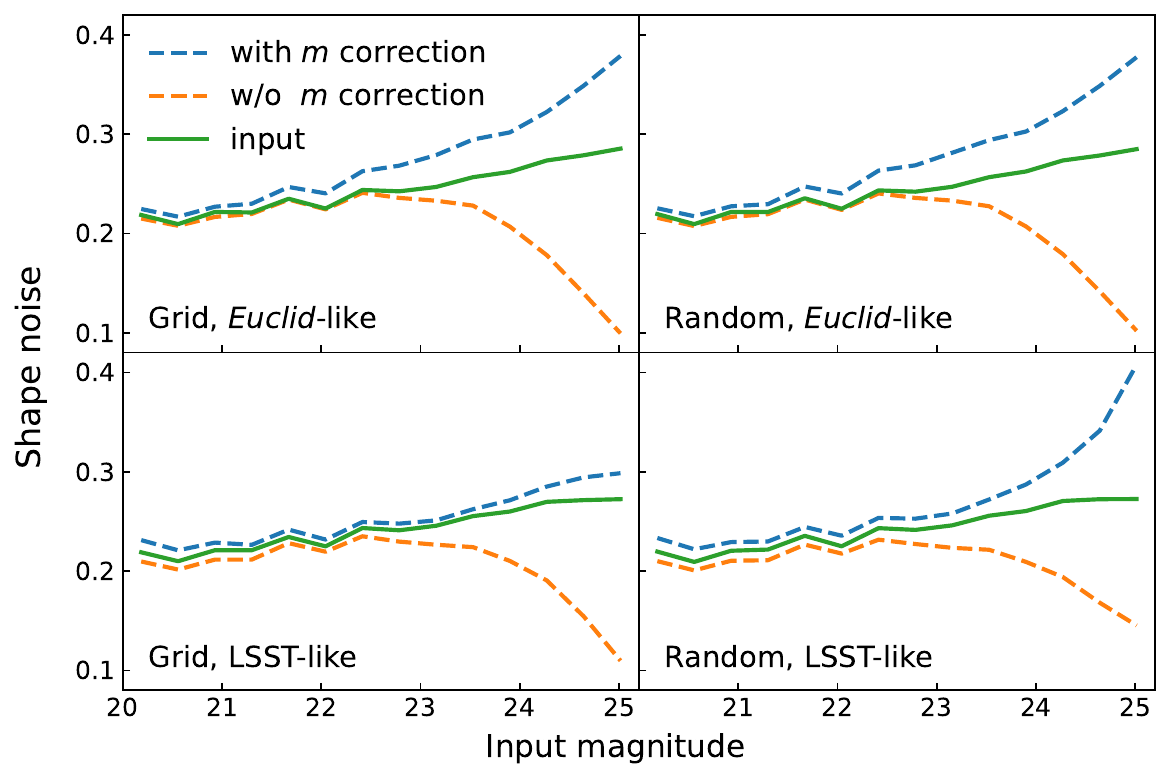}  
\caption{Empirically estimated shape noise, $\sigma_{\rm tot}$ (Eq.~\ref{eq:m-bias_sn}), with (dashed blue lines) and without (dashed orange lines) multiplicative bias correction, compared to the detection bias–corrected intrinsic shape noise (solid green lines), $\sigma_{\rm int}$ (Eq.~\ref{eq:d-bias_input_sn}), as a function of input magnitude. Upper panels: Comparisons for \textit{Euclid}-like setups. The difference between the grid-based and randomly placed configurations is negligible; however, in both cases, applying the bias correction slightly overestimates the shape noise likely arises from second-order effects that are not captured in the multiplicative bias correction. Lower panels: Comparisons for LSST-like setups. For randomly placed galaxies, the corrected $\sigma_{\rm tot}$ rises significantly for magnitudes $\gtrsim 23.5$ relative to the grid case, indicating that blending effects become increasingly important at fainter magnitudes.}
\label{Fig: sn_vs_input}
\end{figure*}

\begin{figure}
\centering
\includegraphics[width=\hsize]{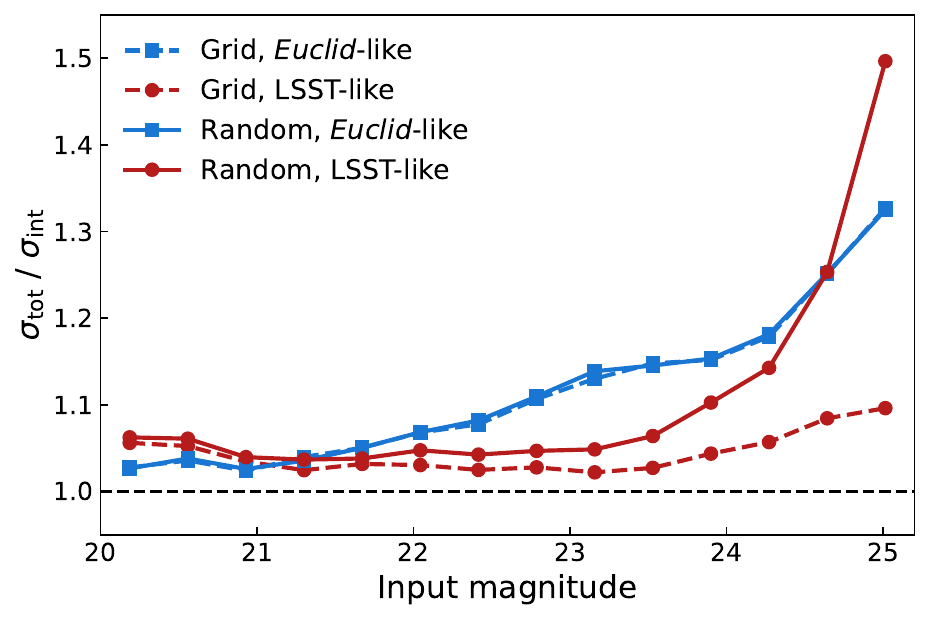}
\caption{Ratios between $\sigma_{\rm tot}$ (Eq.~\ref{eq:m-bias_sn}) and $\sigma_{\rm int}$ (Eq.~\ref{eq:d-bias_input_sn}) as a function of input magnitude. The measured shape noise $\sigma_{\rm tot}$ includes corrections for both multiplicative and additive biases, while the intrinsic shape noise $\sigma_{\rm int}$ is corrected for detection bias (see text for details). For the \textit{Euclid}-like survey, the ratios remain consistent between the grid-based and randomly placed galaxy configurations. In contrast, the increasing discrepancy between the LSST-like setups highlights the growing impact of blending at fainter magnitudes.}
\label{Fig: sn}
\end{figure}

In principle, distortions in observed galaxy shapes provide an unbiased estimate of the lensing signal. In practice, however, lensing-induced shape distortions are much smaller than the intrinsic ellipticities of galaxies. Consequently, the lensing signal must be extracted statistically by averaging over large ensembles of source galaxies. The dominant source of uncertainty in this process is the dispersion in intrinsic galaxy ellipticities, commonly referred to as shape noise. Accurately estimating this shape noise is therefore crucial for quantifying the statistical power of a weak lensing survey.

In practice, we can estimate the shape noise from the total ellipticity dispersion, corrected for shear calibration biases, as follows:
\begin{equation}
\label{eq:m-bias_sn}
    \sigma_{\rm tot} = \frac{1}{2} \left(\sqrt{\frac{\sum_i{w_i^2(e_{i, 1}-c_{1})^2}}{(1+m_1)^2\sum_i w_i^2}} + \sqrt{\frac{\sum_i{w_i^2(e_{i, 2}-c_{2})^2}}{(1+m_2)^2\sum_i w_i^2}}~\right)~,
\end{equation}
where $w_i$ is the weight defined in Eq.~(\ref{eq:weight}), $e_{i,1}$ and $e_{i,2}$ are the two ellipticity components measured by \textsc{ngmix}, and $m_i$, $c_i$ are the multiplicative and additive biases, respectively, which we estimate as a function of input magnitude in bins of width 0.25 mag. The sum runs over all measured galaxies in a given sample.

This estimate reflects the total uncertainty arising from intrinsic galaxy shapes, ellipticity measurement uncertainty, and shear biases. The dashed blue lines in Fig.~\ref{Fig: sn_vs_input} show $\sigma_{\rm tot}$ as a function of input magnitude for various configurations. To highlight the importance of correcting for multiplicative bias, the dashed orange lines show the result without bias correction—that is, by setting $m_i = 0$ in Eq.~(\ref{eq:m-bias_sn}). As seen, neglecting this correction leads to a significant underestimation of the shape dispersion, especially for faint galaxies.

For simulations, we can also compute the intrinsic shape noise under the assumption of perfect ellipticity measurements. We define the detection bias–corrected intrinsic shape noise as
\begin{equation}
\label{eq:d-bias_input_sn}
   \sigma_{\rm int} = \frac{1}{2} \left({\frac{\sqrt{\left<e_{\rm in, 1}^2\right>}}{1+m_{\rm det, 1}}} + {\frac{\sqrt{\left<e_{\rm in, 2}^2\right>}}{1+m_{\rm det, 2}}}\right)\,,
\end{equation}
where $e_{\rm in}$ refers to the input ellipticities, $m_{\rm det}$ is the multiplicative detection bias as a function of magnitude. We omit additive bias here, as it is negligible in our setup. The results are shown as solid green lines in Fig.~\ref{Fig: sn_vs_input}. The use of realistic galaxy morphologies introduces a slight increase in dispersion with magnitude. 

At the bright end, the empirically estimated shape noise, $\sigma_{\rm tot}$ (Eq.~\ref{eq:m-bias_sn}), is slightly larger than the intrinsic shape noise, $\sigma_{\rm int}$ (Eq.~\ref{eq:d-bias_input_sn}). This discrepancy likely arises from second-order effects that are not captured in the multiplicative bias correction. Specifically, we apply a single bias correction per magnitude bin, whereas in reality, the multiplicative bias also depends on other galaxy properties, such as ellipticity. Neglecting these correlations results in a slightly biased estimate of the shape noise. In contrast, $\sigma_{\rm int}$, which is computed directly from input ellipticities, remains unaffected, as detection bias is small overall.

While this second-order effect has no impact on our comparison—since we can use the unbiased $\sigma_{\rm int}$ estimate—it has important implications for real data analyses. Correlations between galaxy properties and multiplicative bias can further amplify the limitation of linear bias correction schemes. If not properly accounted for, these effects can lead to overestimates of the shape noise and, ultimately, impact the accuracy of the covariance matrix used for cosmological parameter inference.

To better visualise the impact of blending on shape noise estimation, Fig.~\ref{Fig: sn} shows the ratio $\sigma_{\rm tot} / \sigma_{\rm int}$ as a function of input magnitude. For relatively bright sources, intrinsic shape noise dominates the total uncertainty, leading to close agreement between $\sigma_{\rm tot}$ and $\sigma_{\rm int}$. The two \textit{Euclid}-like configurations—grid-based (dashed blue line) and randomly placed galaxies (solid blue line)—yield similar results, demonstrating the advantage of a sharper PSF in suppressing blending effects.

In contrast, LSST-like simulations show more significant differences between random (solid red line) and grid-based (dashed red line) setups. When galaxies are placed randomly, the estimated $\sigma_{\rm tot}$ deviates significantly from $\sigma_{\rm int}$, especially for faint galaxies. In the blending-free scenario, this discrepancy is much smaller across the full magnitude range. This demonstrates that blending strongly biases shape noise estimates for LSST-like surveys. For galaxies fainter than magnitude 24.5, the severe blending in the LSST-like setup offsets its greater depth, resulting in worse performance compared to a \textit{Euclid}-like survey with a sharper PSF, but shallower depth.

This underscores the importance of properly accounting for blending when estimating shape noise from measured ellipticities. In our study, where the true input ellipticities are known, we adopt the detection bias–corrected shape noise, $\sigma_{\rm int}$, in subsequent analyses to avoid estimation biases introduced by blending.

\section{Measurement synergy and precision metric}
\label{Sec:ShearComp}

To explore the synergy of combining shape measurements from \textit{Euclid}-like and LSST-like surveys, we require a metric that is relatively simple for analytical study, but also relevant to real-world weak lensing analyses. Various approaches to combining data are still under active investigation \citep{Guy2022zndo...5836022G}. As a first step, we focus on the most salient aspects that can be achieved by combining shape measurements at the catalogue level, and leave the more complicated pixel-level combination for future studies. We describe the methodology for combining shape measurements in Sect.~\ref{Sec:ShapeComb}, and introduce the effective number density in Sect.~\ref{Sec:ShearNeff} as a metric to quantify the overall constraining power of the survey.

\subsection{Synergy between \textit{Euclid}-like and LSST-like surveys}
\label{Sec:ShapeComb}

Following the approach of \citet{Schuhmann2019arXiv190108586S}, we calculate a weighted average of ellipticity measurements for each galaxy detected in both surveys using
\begin{equation}
\label{eq4}
{\bm \epsilon}^{\rm Comb} = \frac{\left(\sigma_{\epsilon}^{\rm LSST}\right)^{-2}{\bm \epsilon}^{\rm LSST} + \left(\sigma_{\epsilon}^{Euclid}\right)^{-2}{\bm \epsilon}^{Euclid}}{\left(\sigma_{\epsilon}^{\rm LSST}\right)^{-2} + \left(\sigma_{\epsilon}^{Euclid}\right)^{-2}}~,
\end{equation}
where $\sigma_{\epsilon}$ refers to the multiplicative bias-corrected uncertainty in ellipticity measurements, as defined in Eq.~(\ref{eq:uncertainty_corrected}). The corresponding uncertainty in the combined measurement is given by
\begin{equation}
\label{eq:sigmaComb}
    \sigma_{\epsilon}^{\rm Comb} = \frac{\sigma_{\epsilon}^{\rm LSST}\sigma_{\epsilon}^{Euclid}}{\sqrt{\left(\sigma_{\epsilon}^{\rm LSST}\right)^{2}+\left(\sigma_{\epsilon}^{Euclid}\right)^{2}}}~,
\end{equation}
which assumes that the measurement uncertainties from the two surveys are independent\textemdash{}as opposed to the intrinsic shapes.

For sources detected in only one of the surveys, we set the uncertainty of the undetected measurement to infinity, effectively excluding it from the combination. Throughout, we refer to the `joint' catalogue as the subset of galaxies detected in both surveys, and the `combined' catalogue as the full set of sources detected in at least one survey.

\subsection{Effective number density}
\label{Sec:ShearNeff}

To first order, the statistical precision of a weak lensing survey is determined by the number of source galaxies and their associated shape noise. Specifically, the uncertainty in the lensing power spectrum scales as $f_{\rm sky}^{-0.5}\sigma_{\rm int}^2n_{\rm eff}^{-1}$~\citep{albrecht2006report,Amara2008MNRAS.391..228A,Chang2013MNRAS4342121C}, where $f_{\rm sky}$ is the fraction of sky covered, and $\sigma_{\rm int}$ is the detection bias–corrected intrinsic shape noise as defined in Eq.~(\ref{eq:d-bias_input_sn}). The effective number density $n_{\rm eff}$ is defined as
\begin{equation}
\label{eq:neff}
n_{\rm eff} = \frac{1}{\Omega}\sum_i\frac{\sigma^2_{\rm int}}{\sigma^2_{\rm int} + \sigma^2_{\epsilon, i}},
\end{equation}
where $\Omega$ is the survey area, and $\sigma_{\epsilon, i}$ is the measurement uncertainty for the $i$-th galaxy, corrected for multiplicative bias as defined in Eq.~(\ref{eq:uncertainty_corrected}). This expression accounts for the fact that measurement noise and blending increase the total uncertainty in shear estimation, thereby reducing the effective number of galaxies contributing to the signal. Thus, $n_{\rm eff}$ provides a practical and informative metric for comparing the statistical power of different weak lensing surveys.

\section{Results}
\label{Sec:Results}

In this section, we compare the constraining power of \textit{Euclid}-like and LSST-like surveys using catalogue-level combinations of ellipticity measurements. We evaluate their synergy through the effective number density, $n_{\rm eff}$, and examine how blending and observing conditions influence the results. We present results based on randomly placed source configurations, which more closely reflect realistic observational scenarios. The number densities for all configurations are summarised in Table~\ref{tab:neff}. 

\begin{figure}
\centering
\includegraphics[width=\hsize]{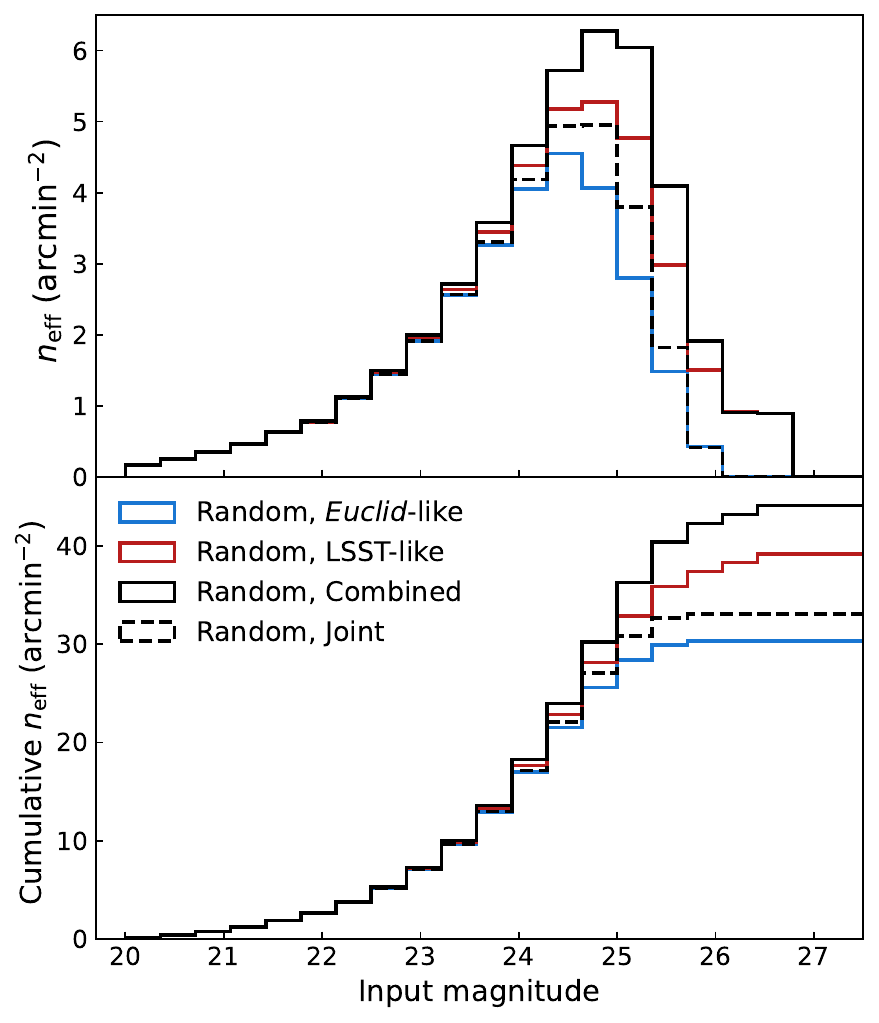}
\caption{Upper panel: Effective number densities as a function of input $r$-band magnitude for \textit{Euclid}-like (blue), LSST-like (red), the joint catalogue (dashed black), and the combined catalogue (solid black). Lower panel: Corresponding cumulative $n_{\rm eff}$ values.}
\label{Fig: neff_catacomb}
\end{figure}

Figure~\ref{Fig: neff_catacomb} shows the bias-corrected effective number density, $n_{\rm eff}$, for random configurations of \textit{Euclid}-like (blue), LSST-like (red), and their combination. The combined catalogue (solid black line) includes all sources detected in either survey, while the joint catalogue (dashed black line) includes only sources detected in both. The input magnitude range is extended to [20.0, 27.5] to account for blending from faint, undetected sources, and better approximate realistic observational conditions. The upper panel displays $n_{\rm eff}$ as a function of magnitude, and the lower panel shows the cumulative $n_{\rm eff}$.

We find that the LSST-like survey consistently achieves a higher $n_{\rm eff}$ than the \textit{Euclid}-like survey, primarily due to its deeper imaging and larger number of detected sources. The combined catalogue (solid black line) further increases $n_{\rm eff}$, particularly at the faint end ($r{\gtrsim}26$ mag), where many sources are detected only in the LSST-like data. In contrast, the contribution from the \textit{Euclid}-like survey is limited by its shallower depth, despite its advantage of a sharper PSF and reduced blending. Overall, the cumulative effective number density of the combined catalogue reaches $44.08~\rm arcmin^{-2}$ over the magnitude range [20.0, 27.5], compared to $39.17~\rm arcmin^{-2}$ for LSST only, and $30.31~\rm arcmin^{-2}$ for \textit{Euclid} only (see Table~\ref{tab:neff} for a summary of all configurations).

For the joint catalogue, which includes only sources detected in both surveys, we find a modest improvement in $n_{\rm eff}$ over the \textit{Euclid}-only case for $r{\gtrsim}23.5$ mag. However, it falls below the LSST-only result at the faint end, as it excludes additional faint sources detected only in the LSST-like images. This underscores that while cross-matching leverages the complementary strengths of both surveys, the full benefit is only realised when all available detections are included in a combined catalogue, rather than limiting the analysis to the subset detected by both. 

However, it is important to note that this conclusion may not directly apply to pixel-level combinations. In a pixel-level joint analysis, a simultaneous fit to exposures from both surveys can be performed when estimating individual ellipticities. Such an approach has the potential to optimally combine the depth and higher S/N of LSST-like imaging with the superior resolution of \textit{Euclid}-like observations. These benefits cannot be fully captured through the catalogue-level combination explored here. Therefore, a full pixel-level joint analysis remains a promising direction, and is warranted for future studies.

\begin{table*}[]
\centering
\caption{Number densities for joint and combined source catalogues.}
\label{tab:neff}
\resizebox{\textwidth}{!}{
\renewcommand{\arraystretch}{1.3}
\begin{tabular}{cccc|ccc|ccc}
\hline
\hline
\multicolumn{10}{l}{Number density ($\rm  arcmin^{-2}$) for joint source catalogue}                               \\ \hline
        & \multicolumn{3}{c|}{Grid {[}20.0, 25.2{]}} & \multicolumn{3}{c|}{Random {[}20.0, 25.2{]}} & \multicolumn{3}{c}{Random {[}20.0, 27.5{]}} \\
Surveys & $Euclid$-like     & LSST-like     & Joint       & $Euclid$-like      & LSST-like      & Joint        & $Euclid$-like      & LSST-like      & Joint        \\ \hline
$n_{\rm input}$             & 38.39 & 38.39 & 38.39 & 38.39 & 38.39 & 38.39 & 186.67 & 186.67 & 186.67 \\
$n_{\rm det}$               & 33.95 & 33.95 & 33.95 & 32.25 & 32.25 & 32.25 & 40.54 & 40.54 & 40.54  \\
$n_{\rm eff}$               & 27.49   & 27.92  & 30.50  & 26.49   & 27.49  & 29.44  & 29.14   & 29.41   & 33.08  \\
$n_{\rm eff} / n_{\rm det}$ & 0.81  & 0.82 & 0.90 & 0.82  & 0.85 & 0.91 & 0.72  & 0.72  & 0.82 \\ \hline
\multicolumn{10}{l}{Number density ($\rm arcmin^{-2}$) for combined source catalogue}                            \\ \hline
Surveys & $Euclid$-like     & LSST-like     & Combined    & $Euclid$-like      & LSST-like      & Combined     & $Euclid$-like      & LSST-like      & Combined     \\ \hline
$n_{\rm det}$               & 34.00   & 38.27  & 38.32  & 33.38   & 36.35  & 37.48  & 42.43   & 93.03   & 94.92  \\
$n_{\rm eff}$               & 27.52   & 31.47  & 34.08  & 27.39   & 30.95  & 33.79  & 30.31   & 39.17   & 44.08  \\
$n_{\rm eff} / n_{\rm det}$ & 0.81  & 0.82 & 0.89 & 0.82  & 0.85 & 0.90 & 0.71  & 0.42  & 0.46 \\ \hline
\end{tabular}
}
\tablefoot{$n_{\mathrm{input}}$ is the input source density;  $n_{\mathrm{det}}$ is the detection number density;  $n_{\mathrm{eff}}$ is the effective number density.
The `joint' catalogue is the subset of galaxies detected in both surveys; `Combined' catalogue is the union of the sources detected in both surveys.} 
\end{table*}

To further investigate the impact of blending in LSST-like surveys, we explore how seeing conditions affect shear measurements (see Appendix~\ref{sec6.2}). In particular, we assess whether selecting exposures with better seeing can mitigate blending-induced biases, with the trade-offs introduced by increased background noise due to reduced exposure stacking. The results show that while selecting exposures with better seeing improves detection performance, it does not fully offset the loss in depth from fewer exposures.

\section{Conclusion}
\label{Sec:Conclusion}

In this paper, we explored the benefits of combining \textit{Euclid}-like and LSST-like surveys at the catalogue level and studied the impact of blending on shear measurements. To this end, we generated 50 realisations of LSST-like and \textit{Euclid}-like images using the \texttt{MultiBand\_ImSim} pipeline, in both grid-based (blending-free) and random (blending-present) configurations. Galaxy shapes were measured using \textsc{ngmix} on objects detected by \textsc{SExtractor}.

By comparing the grid and random setups, we demonstrated that blending significantly biases shear measurements, particularly in LSST-like data due to its broader PSF. This is already evident in the detection rates: while the \textit{Euclid}-like survey shows comparable detection rates between the grid and random configurations, with only a slight decline at the faint end near its magnitude limit, the LSST-like survey exhibits a noticeable drop in detection efficiency when galaxies are randomly placed—even well above its nominal detection limit (see Fig.~\ref{Fig: detec_ratio}).

Blending further affects photometric and shape measurements. In particular, \texttt{MAG\_AUTO} estimates from \textsc{SExtractor} tend to overestimate the brightness of faint objects in the random configuration due to flux contamination from neighbouring sources (Fig.~\ref{Fig: mag_corr}). The influence of blending is further reflected in the measured multiplicative biases (Fig.~\ref{Fig: m1_4cases}), which propagate into shape noise estimates (Figs.~\ref{Fig: sn_vs_input} and \ref{Fig: sn}), complicating their applications.

These results highlight the importance of incorporating realistic blending effects into image simulations when evaluating survey performance, and assessing the potential synergy between ground- and space-based weak lensing data. Accounting for these effects, we find that the most effective catalogue-level synergy is achieved by combining all galaxies detected in either survey—referred to as the combined catalogue in our analysis. This approach yields an effective galaxy number density of $44.08~\rm arcmin^{-2}$ over the magnitude range of 20.0 to 27.5, compared to $39.17~\rm arcmin^{-2}$ for LSST-like data alone, and $30.31~\rm arcmin^{-2}$ for \textit{Euclid}-like data alone. These results demonstrate that catalogue-level synergy between deep ground-based and high-resolution space-based surveys can enhance the statistical power of weak lensing measurements by combining complementary information.

Restricting the analysis to overlapping sources detected in both surveys yields a more modest improvement of ${\sim}12\%$ compared to either survey alone. Notably, the LSST-like and \textit{Euclid}-like surveys achieve comparable effective number densities for the overlapping sample (see Table~\ref{tab:neff} for detailed values), reflecting a compensating performance in galaxy shape measurement precision between the greater depth of LSST-like data and the higher resolution of \textit{Euclid}-like data. However, fully realising the potential of joint-object analyses requires a pixel-level approach, where joint modelling during the shape measurement process can more effectively exploit the complementary strengths of both data sets.

This has been partially explored by \citet{Schuhmann2019arXiv190108586S}, who found that pixel-level joint modelling yields only a modest improvement of ${\sim}5\%$ in effective number density for faint sources compared to catalogue-level combinations. However, their simulations did not include blending effects, which—as demonstrated in this work—can have a significant impact on shape measurement. Their results suggest that for non-blended sources, the more accessible catalogue-level combination is probably sufficient for the shear measurement synergy. However, we emphasise that catalogue-level combination, even with first-order multiplicative biases correction, does not by itself mitigate all survey-specific systematics. Residual uncertainties in PSF modelling, selection functions, photometric calibration, and redshift distributions can differ between LSST-like and Euclid-like data, and may propagate into cosmological inference if not properly accounted for. Robust mitigation of these survey-specific systematics will therefore require further development of unified calibration pipelines. In particular, given the depth of LSST-like surveys and the associated high blending fraction, a thorough investigation of pixel-level synergies that incorporates realistic blending is still warranted. Recent studies \citep{congedo2024A&A...691A.319E} injected galaxies at their true sky positions, preserving both clustering and intrinsic property correlations, which allows them to isolate blending bias in a more realistic context. Additionally, joint‐fitting methods such as lensMC model all neighbours simultaneously, offering a promising deblending alternative. Exploring these approaches will be valuable extensions of our current catalogue‐level analysis.

Given the significant blending effects in LSST-like data, we further investigated whether selecting subsets of exposures with better seeing could help mitigate these impacts. We compared shear measurements for galaxies selected under different seeing conditions in the LSST-like survey: all exposures, the best-half, and the best-quarter in terms of seeing quality, while accounting for the increased background noise associated with reduced stacking depth. Our results show that selecting better-seeing exposures does not compensate for the loss in S/N caused by fewer stacked images. The most effective mitigation of blending only arises from the sharper PSF provided by space-based observations, such as those from a \textit{Euclid}-like survey. These findings reflect the limitations imposed by atmospheric turbulence on ground-based data, and highlight the advantages of space-based imaging for reliable shape measurement in weak lensing studies.

\begin{acknowledgements}
We gratefully acknowledge the anonymous referee for their
valuable comments that helped us to improve the manuscript. SZ acknowledges the support from the Deutsche Forschungsgemeinschaft (DFG) SFB1491. SSL acknowledges funding from the programme "Netzwerke 2021", an initiative of the Ministry of Culture and Science of the State of Northrhine Westphalia. HHo and SSL acknowledge support from the European Research Council (ERC) under the European Union’s Horizon 2020 research and innovation program with Grant agreement No. 101053992. 
\end{acknowledgements}

  \bibliographystyle{aa.bst} 
  \bibliography{aa.bib}

\begin{appendix} 

\section{Different seeings in LSST-like images}
\label{sec6.2}

Given the significant impact of blending in LSST-like surveys due to their larger PSFs, we investigate whether selecting exposures with better seeing can mitigate its effect on shear measurement. However, using only a subset of exposures comes at the cost of increased background noise due to reduced stacking depth, making the trade-off non-trivial.

Following the procedure described in Sect.~\ref{Sec:ImaSim}, we simulate LSST-like observations using subsets with better seeing conditions. We select fields from different percentiles of the seeing distribution to account for variability due to changing atmospheric conditions.
To do so, we randomly select 30 fields—covering a total area of $60~\rm deg^2$—from the best half and best quarter of seeing values in the full KiDS DR4 sample. To account for the reduced number of stacked exposures, we increase the background noise by factors of $\sqrt{2}$ and 2, respectively. The resulting average seeing values are $0\farcs611 \pm 0\farcs051$ and $0\farcs570 \pm 0\farcs034$, with effective background depths of 27.13 and 26.75 $\rm mag~arcsec^{-2}$. These are still deeper than the \textit{Euclid}-like images, which have a background depth of 24.71 $\rm mag~arcsec^{-2}$.

\begin{figure}
  \centering
  \includegraphics[width=\columnwidth]{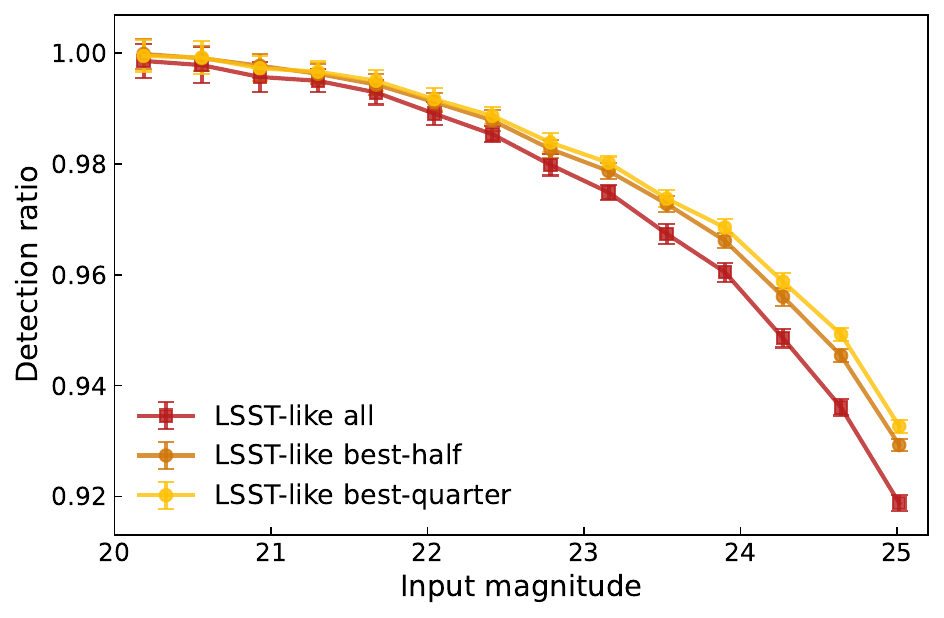}
  \caption{Detection efficiency relative to the LSST-like grid-based case with all stacked exposures as a function of input magnitude, shown for three subsets: all exposures (red), best-half seeing (orange), and best-quarter seeing (yellow).}
  \label{Fig: det_ratio_seeing}
\end{figure}

To validate the impact of seeing selection on source detection, we compare the detection efficiency of three LSST-like subsets (using all exposures, best-half seeing, and best-quarter seeing) relative to our grid-baseline case constructed from all exposures. Figure~\ref{Fig: det_ratio_seeing} shows that the best-half (orange) and best-quarter (yellow) seeing subsets yield slightly improved detection efficiencies compared to the full-stack subset (red). These modest gains illustrate that the advantage of reduced blending under better seeing conditions is partially offset by increased background noise arising from fewer exposures.

\begin{figure}
\centering
\includegraphics[width=\hsize]{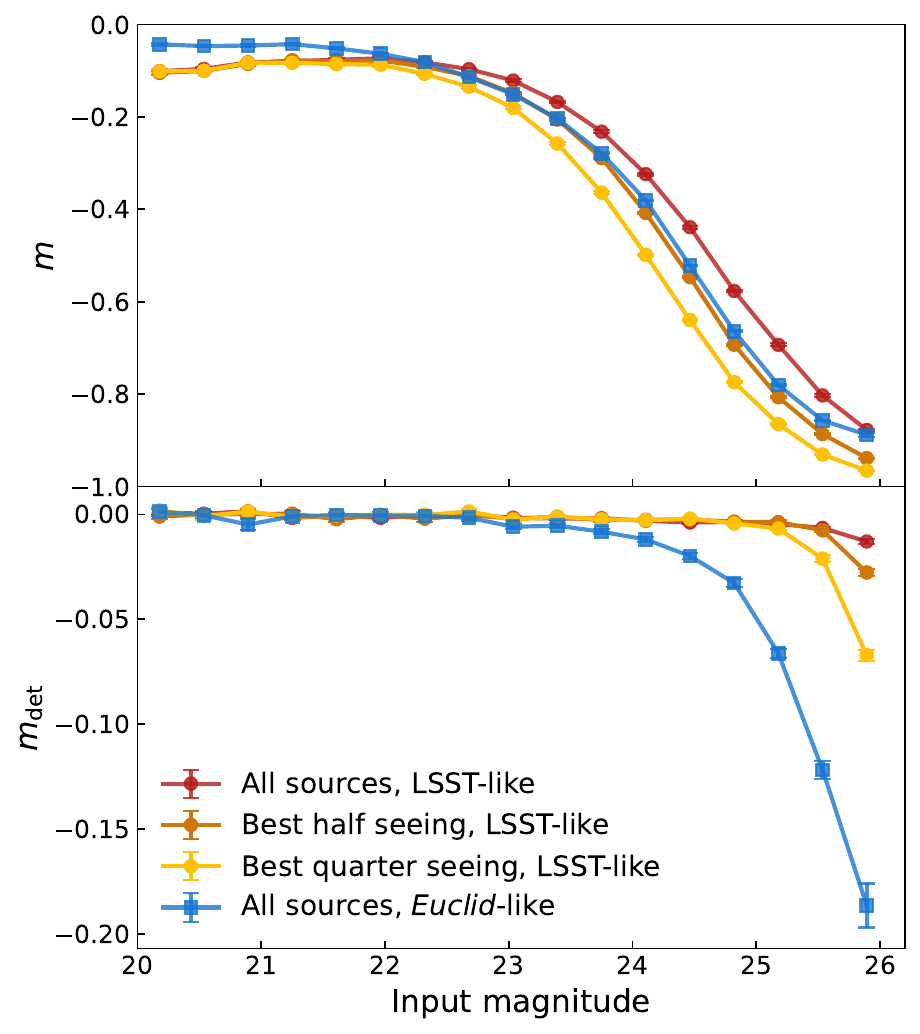}
\caption{Upper panel: multiplicative bias, defined as $m\equiv(m_1+m_2)/2$ as a function of the input magnitude of galaxies for \textit{Euclid}-like and different selections of seeings in LSST-like surveys. Lower panel: multiplicative detection bias $m_{\rm det}$ with the input magnitude with galaxies are placed randomly, also defined as $m_{\rm det}\equiv(m_{\rm det, 1}+m_{\rm det, 2})/2$.}
\label{Fig: m_seeings}
\end{figure}

Figure~\ref{Fig: m_seeings} shows the multiplicative bias (upper panel) and detection bias (lower panel) as functions of input magnitude for LSST-like surveys under three different seeing selections. Selecting galaxies from the best-half or best-quarter seeing conditions results in larger multiplicative biases compared to using the full set of exposures. This suggests that the improved overall seeing condition does not compensate for the increased background noise introduced by reducing the number of stacked images.

\begin{figure}
\centering
\includegraphics[width=\hsize]{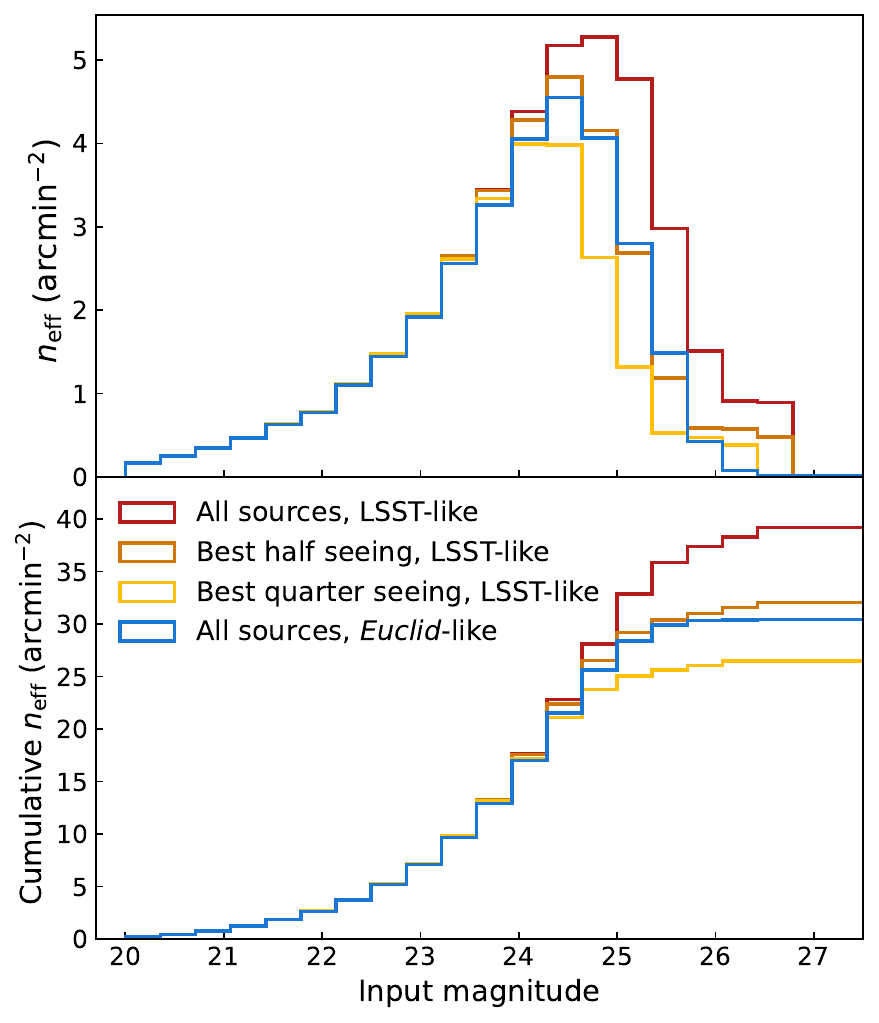}
\caption{The effective number densities (upper panel) and cumulative effective number densities (lower panel) for \textit{Euclid}-like and different seeings selections of LSST-like surveys, across an input $r$-band magnitude range of [20.0, 27.5].}
\label{Fig: seeingneff}
\end{figure}

Figure~\ref{Fig: seeingneff} presents the corresponding multiplicative bias–corrected $n_{\rm eff}$ and cumulative $n_{\rm eff}$. Consistent with the bias results in Fig.~\ref{Fig: m_seeings}, selecting only the best-half seeing exposures yields lower $n_{\rm eff}$ than using all exposures, particularly for magnitudes $r{\gtrsim}24.5$. Further restricting to the best-quarter seeing leads to an even greater reduction in $n_{\rm eff}$. These results indicate that, for an LSST-like ground-based survey, mitigating blending by selecting only the best-seeing exposures is not an effective strategy when it comes at the cost of increased noise from reduced stacking depth.

Nevertheless, it is noteworthy that the $n_{\rm eff}$ of the best-half seeing subset is comparable to that of the \textit{Euclid}-like survey, despite the latter having a significantly shallower depth (by approximately $2.5~\rm mag~arcsec^{-2}$). This illustrates the high efficiency of the sharper space-based PSF in recovering reliable shape measurements. It also implies how much sharper the PSF needs to be in order to compensate for shallower observations.

\end{appendix}

\end{document}